\documentclass[aps,superscriptaddress,showpacs,preprint]{revtex4}
\usepackage{epsfig}
\usepackage{graphicx}
\usepackage{psfrag}
\usepackage{subfigure}
\usepackage{bm}
\newcommand{\be}{\begin{equation}}

\newcommand{\ee}{\end{equation}}
\newcommand{\bea}{\begin{eqnarray}}
\newcommand{\eea}{\end{eqnarray}}

\begin{document}
\hfill$\vcenter{
 \hbox{\bf BARI-TH 498/04}}$
\title{Meissner masses in the gCFL phase of QCD}
\author{R.Casalbuoni}
\affiliation{Dipartimento di Fisica, Universit\`a di Firenze,
I-50125 Firenze, Italia} \affiliation{I.N.F.N., Sezione di
Firenze, I-50125 Firenze, Italia}
\author{R.Gatto}
\affiliation{D\'epart. de Physique Th\'eorique, Universit\'e de
Gen\`eve, CH-1211 Gen\`eve 4, Suisse}
\author{M.Mannarelli}
\affiliation{Cyclotron Institute and Physics Department, Texas
A\&M University, College Station, Texas 77843-3366}

\author{G.Nardulli}
\author{M.Ruggieri}
\affiliation{Universit\`a di Bari, I-70126 Bari, Italia}
\affiliation{I.N.F.N., Sezione di Bari, I-70126 Bari, Italia}
%
%\begin{center}
%{\Large\bf\boldmath {Meissner masses for gluons in the gCFL phase of QCD}}
%\\ \rm \vskip1pc {\large
%R. Casalbuoni$^{a,b}$,  R. Gatto$^c$, M. Mannarelli$^{f}$, G.
%Nardulli$^{d,e}$,  M.Ruggieri$^{d,e}$}\\
%\vspace{5mm} {\it{$^a$Dipartimento di Fisica, Universit\`a di
%Firenze, I-50125 Firenze, Italia
%\\
%$^b$I.N.F.N., Sezione di Firenze, I-50125 Firenze, Italia\\
%$^c$D\'epart. de Physique Th\'eorique, Universit\'e de Gen\`eve,
%CH-1211 Gen\`eve 4, Suisse\\ $^d$Dipartimento di Fisica,
%Universit\`a di Bari, I-70126 Bari, Italia  \\$^e$I.N.F.N., Sezione
%di Bari, I-70126 Bari, Italia \\$^f$ Cyclotron Institute and Physics
%Department, Texas A\&M University, College Station, Texas
%77843-3366}}
%\end{center}
%%% ----------------------------------------------------------------------

\begin{abstract}
We calculate the Meissner masses of gluons in neutral three-flavor color superconducting matter for finite strange quark mass. In the CFL phase the
Meissner masses are slowly varying function of the strange quark mass. For large strange quark mass, in the so called gCFL phase, the Meissner masses
of gluons with colors $a=1,2,3$ and $8$ become imaginary, indicating an instability.
\end{abstract}
 \maketitle

\section{Introduction}
At asymptotic densities cold quark matter is in the Color Flavor
Locked (CFL) phase of QCD \cite{Alford:1998mk} (see also
\cite{Rajagopal:2000wf} and \cite{Rapp:1998zu}). This state is
characterized by nine gapped fermionic quasi-particles (3$\times
3$, for color and flavor) and by electric neutrality even for non
vanishing quark masses $M_j$, provided $M_j\neq 0$ does not
destroy the CFL phase \cite{Rajagopal:2000ff}. For lower
densities, it has recently been shown
\cite{Alford:2003fq,Alford:2004hz,Fukushima:2004zq} that,
including the strange quark mass $M_s$, requiring electrical and
color neutrality, and imposing weak equilibrium, a phase
transition occurs, from the CFL phase to a new phase, called
gapless CFL or gCFL. In the gCFL phase  seven fermionic
quasiparticles have a gap in the dispersion law,  but the
remaining two  are gapless. At zero temperature the transition
from the CFL to the gCFL phase takes place at $M_s^2/\mu_b \approx
2 \Delta$, where $\mu_b$ is the quark chemical potential and
$\Delta$ is the gap parameter. At non zero temperature the
situation is more involved \cite{Ruster:2004eg,Fukushima:2004zq}
and also the existence of mixed phases \cite{Reddy:2004my} has to
be taken into account. The next phase at still lower densities is
difficult to determine and the crystalline color superconductive
phase is a candidate \cite{Casalbuoni:2004wm}.

 The aim of this paper is to investigate the
 dependence of Meissner masses on the
strange quark mass in the gCFL phase. For two flavors a similar
analysis  has recently been performed  by Huang and Shovkovy
 \cite{Huang:2004bg}. Imposing weak equilibrium and neutrality
 they compute the gluon Meissner mass in the 2SC phase and show
that an instability arises in a certain range of values of the
parameters, with some gluon masses becoming imaginary. We present
numerical evidence that a similar instability is also present in
the gCFL phase. By the same method we also investigate the
dependence of Meissner masses on $M_s$ in the gapped phase (CFL
with $M_s\neq 0$). Our calculational scheme is the  High Density
Effective Theory (HDET) \cite{Hong:1998tn,Nardulli:2002ma}, which
allows a significant reduction of the computational complexity.

The plan of this paper is as follows. In Section \ref{HDET} we
derive the effective Lagrangian and the grand potential in the
HDET scheme. In Section \ref{polarizationsection} we determine the
polarization tensor for the gluons in the HDET approximation. In
Section \ref{numeric} we present and discuss the numerical results
for the Meissner masses as a function of the strange quark mass.
The conclusions are summarized in Section \ref{conclusions}.

\section{HDET approach to the gCFL phase \label{HDET}}
Following Ref. \cite{Alford:2004hz}  the Lagrangian for gluons and
ungapped quarks with $M_u=M_d=0$ and $M_s\neq0$ can be written as
follows (color, flavor and spin indices suppressed):
\begin{equation}
{\cal L}=\bar{\psi}\,\left(i\,D\!\!\! / -{\bf M}+ \bf{\mu}
\,\gamma_0\right)\,\psi \label{lagr1}
\end{equation}
where ${\bf M} = {\rm diag}(0,0,M_s) $ is the mass matrix  in
flavor space and the matrix of chemical potential is given by
\cite{Alford:2003fq} \be {\bf\mu}_{ij}^{\alpha\beta} = \left(\mu_b
\delta_{ij}- \mu_Q Q_{ij}\right)\delta^{\alpha\beta}+ \delta_{ij}
\left(\mu_3 T_3^{\alpha\beta}+\frac{2}{\sqrt 3}\mu_8
T_8^{\alpha\beta}\right) \ee ($i,j =1,3 $ flavor indices;
$\alpha,\beta =1,3 $ color indices). Moreover $T_3 = \frac 1 2
{\rm diag}(1,-1,0)$, $T_8 = \frac{1}{2 \sqrt 3 }{\rm
diag}(1,1,-2)$ in color space and $Q= {\rm diag} (2/3,-1/3,-1/3)$
in flavor space;  $ \mu_Q$ is the electrostatic chemical
potential; $\mu_3, \mu _8$ are the color chemical potentials
associated respectively to the color charges $T_3$ and $T_8$;
$\mu_b$ is quark chemical potential which we fix to $500$ MeV. As
usual in the HDET, to get rid of the Dirac structure we introduce
velocity dependent fields of positive (negative) energy $\psi_{\bf
v} (\Psi_{\bf v}$) by the Fourier decomposition
\cite{Nardulli:2002ma}
\begin{equation}
\psi=\sum_{\bf v}e^{i\,\mu_b\,{\bf v}\cdot{\bf
x}}\,\left(\psi_{\bf v}+\Psi_{\bf v}\right)\label{decomp} \,,
\end{equation}
where ${\bf v}$ is a unit vector representing the Fermi velocity
of the quarks. Substituting the expression (\ref{decomp}) in the
Eq.(\ref{lagr1}) at the leading order in $M_s^2/\mu_b$ we obtain
the HDET Lagrangian
\begin{equation} {\cal L}= \sum_{\bf v}\, \psi_{\bf v}^\dagger\left(i V \cdot D + {\bf \delta\mu} - \frac{M^2}{ 2 \mu_b } \right) \psi_{\bf
v} - P_{\mu\nu} \psi_{\bf v}^\dagger\left( \frac{D_{\mu} D_{\nu}}{
i \tilde V \cdot D + 2 \mu_b }\right) \psi_{\bf v} \label{L11}\, ,
\end{equation}
where $V^\mu=(1,{\bf v}),~\tilde{V}^\mu=(1,-{\bf v})$ and
\begin{equation}
P^{\mu\nu}= g^{\mu \nu} - \frac{1}2 \left(V^{\mu}\tilde V^{\nu}+
\tilde
  V^{\mu} V^{\nu} \right)\, .
\end{equation}It is clear that, at this order of approximation,
the effect of  $M_s\neq 0$ is to reduce the chemical potential of
the strange quarks. Let us now define a new  basis $\psi_A$ for
the spinor fields:
\begin{equation}
\psi_{\alpha i}=\sum_{A=1}^{9}\,\left(F_A\right)_{\alpha
i}\,\psi_A \, ,
\end{equation}
where the matrices $F_A$ can be expressed  by
\begin{eqnarray}
&&F_{1} = \frac{1}{3}I_0+T_3+\frac{1}{\sqrt{3}}T_8, ~~~~~F_{2}=\frac{1}{3}I_0-T_3+\frac{1}{\sqrt{3}}T_8, ~~~~~F_{3}=\frac{1}{3}I_0-\frac{2}{\sqrt{3}}T_8,     \\
&&F_{4,5}= T_1\pm i\,T_2, ~~~~~~~~~~~~~~~ F_{6,7}= T_4\pm
i\,T_5,~~~~~~~~~~~~~~~F_{8,9}= T_6\pm i\,T_7 \, ,
        \label{defFFmatrices}
\end{eqnarray}
with  $T_a=\lambda_a/2$  the  $SU(3)$  generators and $I_0$  the
identiy matrix. Introducing  the Nambu-Gorkov fields
\begin{equation}
\chi_A=\frac{1}{\sqrt{2}}\left(\begin{array}{c}
  \psi_{\bf v} \\
  C\,\psi^*_{- \bf v}
\end{array}\right)_A \label{chi} \label{nambu-gorkov}
\end{equation}
 the  kinetic part of the Lagrangian  (\ref{L11}) reads
\begin{equation}
{\cal L}_0=\sum_{\bf v}\,\chi^\dagger_A\,\left(
\begin{array}{cc}
  \left(V\cdot \ell\, + \delta\mu_{A} - \frac{M^2_A}{2 \mu_b}\right)\delta_{AB} & 0 \\
  0 &  \left(\tilde V\cdot \ell\, - \delta\mu_{A} + \frac{M^2_A}{2 \mu_b}\right)\delta_{AB}
\end{array}
\right)\,\chi_B \label{kinetic}
\end{equation}
where\begin{equation}\delta\mu_{A} \ = \ \left(\delta\mu_{ru},
\delta\mu_{gd}, \delta\mu_{bs}, \delta\mu_{rd}, \delta\mu_{du},
\delta\mu_{rs}, \delta\mu_{bu}, \delta\mu_{gs},
\delta\mu_{bd}\right) \label{chemPotMatr}\end{equation} and
$M^2_A=M^2_s(0,0,1,0,0,1,0,1,0)$. If we define \be\delta\mu_A^{\rm
eff} = \delta\mu_A - \frac{M^2_A}{2 \mu_b}\, \, ,\ee we may recast
Eq. (\ref{kinetic}) as
\begin{equation}
{\cal L}_0=\sum_{\bf v}\,\chi^\dagger_A\,\left(
\begin{array}{cc}
  \left(V\cdot \ell\, + \delta\mu_A^{\rm eff} \right) \delta_{AB} & 0 \\
  0 &  \left( \tilde V\cdot \ell\, -  \delta\mu_A^{\rm eff}\right) \delta_{AB}
\end{array}
\right)\,\chi_B \label{kineticBIS} \, ,
\end{equation}
which is formally equivalent to the Lagrangian for massless quarks
with different chemical potentials.

In the  gapless color-flavor-locking (gCFL) phase the symmetry
breaking is induced by the condensate \cite{Alford:2003fq}
\begin{equation}
\Delta_{ij}^{\alpha\beta}\equiv<\psi_{i\alpha}\,C\,\gamma_5\,\psi_{\beta
j}> = \sum_{I=1}^{3}\,\Delta_I\,\epsilon^{\alpha\beta
I}\,\epsilon_{ijI}\label{cond}
\end{equation}
and the corresponding gap term in the Lagrangian in the mean field
approximation  is
\begin{equation}
{\cal
L}_{\Delta}=-\frac{1}{2}\,\Delta^{\alpha\beta}_{ij}\,\sum_{\bf
v}\,\psi^T_{\alpha i,-{\bf v}}\,C\,\gamma_5\,\psi_{j\beta,+{\bf
v}}\,+\,h.c. \label{gapLagrUNO} \, .
\end{equation}
In the Nambu-Gorkov basis (\ref{nambu-gorkov}) the gap  term reads
\begin{equation}
{\cal L}_{\Delta}= \sum_{\bf v}\,\chi^\dagger_A\,\left(
\begin{array}{cc}
  0 & -\Delta_{AB} \\
  -\Delta_{AB} & 0
\end{array}
\right)\,\chi_B\label{gapLagrDUE} \, ,
\end{equation}
where $\Delta_{AB}$ is   the $9\times 9$ matrix defined  by
\begin{eqnarray}
\Delta_{AB}&=&-\sum_{I=1}^{3}\,\Delta_I\,\text{Tr}\left[F_A^T\,\epsilon_I\,F_B\,\epsilon_I\right] \nonumber \\
&=&\left(
\begin{array}{ccccccccc}
  0 & \Delta_3 & \Delta_2 & 0 & 0 & 0 & 0 & 0 & 0 \\
  \Delta_3 & 0 & \Delta_1 & 0 & 0 & 0 & 0 & 0 & 0 \\
  \Delta_2 & \Delta_1 & 0 & 0 & 0 & 0 & 0 & 0 & 0 \\
  0 & 0 & 0 & 0 & -\Delta_3 & 0 & 0 & 0 & 0 \\
  0 & 0 & 0 & -\Delta_3 & 0 & 0 & 0 & 0 & 0 \\
  0 & 0 & 0 & 0 & 0 & 0 & -\Delta_2 & 0 & 0 \\
  0 & 0 & 0 & 0 & 0 & -\Delta_2 & 0 & 0 & 0 \\
  0 & 0 & 0 & 0 & 0 & 0 & 0 & 0 & -\Delta_1 \\
  0 & 0 & 0 & 0 & 0 & 0 & 0 & -\Delta_1 & 0
\end{array}
\right)\, .\label{gapMatrNEW}
\end{eqnarray}From Eqs.(\ref{kineticBIS}) and (\ref{gapLagrDUE}) one immediately obtains the inverse fermion propagator that in momentum space is given by
\begin{equation}
S^{-1}_{AB}(\ell)=\left(
\begin{array}{cc}
  (V\cdot \ell  + \delta\mu^{eff}_{A})\delta_{AB} & -\Delta_{AB} \\
 -\Delta_{AB} & (\tilde{V}\cdot \ell - \delta\mu^{eff}_{A})\delta_{AB}
\end{array}
\right)  \ .\label{invFerm}
\end{equation}It can be inverted to get the fermion propagator
\begin{equation}
S_{AB}(l)=\left(\begin{array}{cc} \left(P \,
\Delta\,(\tilde{V}\cdot l - \delta\mu_{eff}
)\,\Delta^{-1}\right)_{AB}&
P_{AC}\,\Delta_{CB}\\
D_{AC}\,\Delta_{CB}& \left(D\,\Delta\,(V\cdot l +
\delta\mu_{eff})\,\Delta^{-1}\right)_{AB}
\end{array}\right) \label{FermProp} \,
\end{equation}
where \be P \,=\, \frac{1}{\Delta\,(\tilde{V}\cdot l - \delta\mu_{eff})\,\Delta^{-1}\,(V\cdot l + \delta\mu_{eff})-\Delta^2} \ee and $D=P(V
\leftrightarrow \tilde{V},\delta\mu_{eff}  \leftrightarrow -\delta\mu_{eff})$. From the poles of the propagator we can now determine the dispersion
laws of the quasiparticles. The  knowledge of the dispersion laws allows the evaluation of the grand potential which in the limit of zero temperature
is given by \be \Omega \,=\, -\frac{1}{2 \pi^2}\int_{0}^{\Lambda} dp\,p^2 \sum_{j=1}^{9} |\epsilon_j(\ell_\parallel)| + \frac{1}{G}
(\Delta_1^2+\Delta_2^2+\Delta_3^2)- \frac{\mu_Q^4}{12 \pi^2}\, ,\label{grandpotential}\ee where $\Lambda $ is the ultraviolet cutoff,
$\epsilon_j(\ell_\parallel)$ are the quasi particle dispersion laws, $\ell_\parallel$ is the quark momentum measured from the Fermi surface
($p=\mu_b+\ell_\parallel$), and $G$ is the Nambu-Jona Lasinio coupling constant. We fix $G$ as in \cite{Alford:2003fq}, requiring that the value of
the gap is $25$ Mev for $M_s=0$. In order to enforce electrical and color neutrality one has to minimize the grand potential with respect to $\mu_Q,
\mu_3$ and $\mu_8$. Including the stationary conditions with respect to the gap parameters $\Delta_1, \Delta_2, \Delta_3$ one ends up with a system
of six equations which must be solved simultaneously. Once this system of equations is solved one may express $\Delta_1, \Delta_2, \Delta_3$ and the
chemical potentials $\mu_Q, \mu_3$ and $\mu_8$ as a function of $M_s^2/\mu_b$. We have numerically checked that using the grand potential
(\ref{grandpotential}), with $\mu_b=500~MeV$ and $\Lambda=800~MeV$, we recover the results of Ref. \cite{Alford:2003fq} with an error of  $5\%$.

\section{Polarization tensor of  gluons \label{polarizationsection}}
To compute gluon Meissner masses  we evaluate the polarization
tensor $\Pi_{ab}^{\mu\nu}(p)$.
 In the HDET approach, at the leading
order in $g \mu_b$, there are two contributions to the polarization
tensor:  The self-energy (s.e.) diagram and the tadpole (tad)
diagram (see e.g. Fig. 2 in  \cite{Casalbuoni:2002my}).  To evaluate
 the self-energy diagram  we extract  the trilinear quark-gluon
coupling by the minimal coupling term in the Lagrangian
(\ref{L11}):
\begin{equation}
{\cal L}_1=i\,g\,\sum_{\bf v}\,\psi^\dagger_{i\alpha,{\bf
v}}\,i\,V^\mu\,A_{\mu}^a\,(T_a)^{\alpha\beta}\,\psi_{\beta j,{\bf
v}} \label{L1}
\end{equation}
which can be rewritten in the Nambu-Gorkov  basis
(\ref{nambu-gorkov}) as \be {\cal L}_1=i\,g\,\sum_{\bf
v}\,\chi_A^\dagger\,\left(
\begin{array}{cc}
 i\, V\cdot A_a\, h^a_{AB} & 0 \\
  0 & -i\,\tilde{V}\cdot A_a\, h^{a *}_{AB}
\end{array}
\right)\,\chi_B \equiv i\,g\,\sum_{{\bf
v}}\,\chi_A^\dagger\,\tilde{H}^{a\mu}_{AB}\,\chi_B\,A^a_\mu\label{L1new}
\ee where $ h^a_{AB}=Tr[F^\dagger_A\,T^a\,F_B]$. Therefore the
self-energy contribution to the polarization tensor is given by:
\be i\,\Pi_{ab}^{s.e.\mu\nu}(p) =
\frac{g^2\mu_b^2}{4\,\pi^3}\int\frac{d{\bf v }}{4\,\pi}\, \int d^2
\ell\,Tr\left[S(\ell)\,{\tilde H}^{a\mu}\,S(\ell + p)\,{\tilde
H}^{b\nu}\right] \label{self-energy} \, .\ee  In order to evaluate
the tadpole diagram contribution  we extract the quadrilinear
quark-gluon coupling from the second term on the r.h. side of
Eq.(\ref{L11}) \be {\cal L}_2=-g^2\,\sum_{\bf
v}\,\psi^\dagger_{\bf v}\,\frac{T_a\,T_b}{\tilde{V}\cdot \ell +
2\,\mu_b}\,\psi_{\bf v} \,P_{\mu\nu}\,A^\mu_a\,A^\nu_b
\label{L2LL} \, .\ee In the Nambu-Gorkov basis this term reads
\begin{equation}
{\cal L}_2=-g^2\,\sum_{\bf v}\,\chi_A^\dagger\,\left(
\begin{array}{cc}
  \frac{d_{AB}^{ab}}{\tilde{V}\cdot \ell +
2\,\mu_b} & 0 \\
  0 & \frac{d_{AB}^{ab *}}{V \cdot \ell +
2\,\mu_b}   \end{array}
\right)\,\chi_B\,P_{\mu\nu}\,A^\mu_a\,A^\nu_b   \equiv
-g^2\,\sum_{\bf
v}\,\chi_A^\dagger\,Y_{AB}^{ab}\,\chi_B\,P_{\mu\nu}\,A^\mu_a\,A^\nu_b
\label{L2Lnew} \, ,\end{equation} with $
d^{ab}_{AB}=Tr[F^\dagger_A\,T^a\,T^b\,F_B]$. The tadpole
contribution is then evaluated to be
\begin{equation}
i\,\Pi_{ab}^{tad,\mu\nu} =
-2\,g^2\,\frac{4\,\pi}{16\,\pi^4}\int\frac{d{\bf v
}}{4\,\pi}\,P^{\mu\nu}\,\int d\ell
d\ell_\parallel\,Tr\,\left[i\,S(l)\,\left(\begin{array}{cc}
  (\mu_b+\ell_\parallel)^2 & 0 \\
  0 & (\mu_b-\ell_\parallel)^2
\end{array}\right)\,Y^{ab}\right] \label{tadpole} \, .
\end{equation}
 Finally, in the HDET approximation, the gluon
 polarization tensor is given by:
 \be
 \Pi^{\mu\nu}_{ab}(p)\,=\,  \Pi^{s.e.,\mu\nu}_{ab}(p) +
\Pi^{tad,\mu\nu}_{ab}(p)  \, . \label{polarization} \ee We note
that this result for the polarization tensor is correct at the
order ${\cal O} (M_s/\mu_b)^2$.

\section{Numerical results \label{numeric}}

\begin{figure}[t]
\centering {\includegraphics[width=14cm]{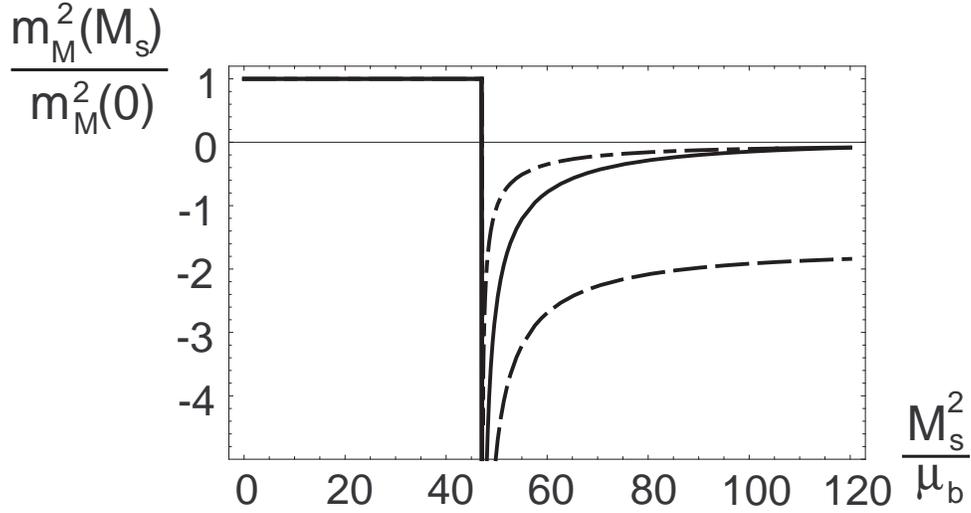}}
\caption{\label{masse12and38}{ \rm Squared values of the  Meissner masses in units of
$m_M^2$ (see Eq. (\ref{meissnerCFL})) as a function of $M_s^2/\mu_b$
(in MeV) for gluons $a=1,2,3,8$. The solid line denotes the gluons with colors
$a=1,2$.  The dashed line denotes the gluons with color $a=3$; finally,
the dot-dashed line is for $a=8$.}}
\end{figure}

The Meissner masses of the gluons are  obtained by the eigenvalues
of the polarization tensor (\ref{polarization}) in the static limit
$p_0=0,{\bf p}\rightarrow 0$. In the CFL phase with $M_s=0$ the
Meissner masses are degenerate and one has
\cite{Son:1999cm,Casalbuoni:2000na} \be m^2_{M}= \frac{\mu_b^2
g^2}{\pi^2}\left(-\frac{11}{36} - \frac{2}{27} \ln 2 + \frac 1 2
\right) \, . \label{meissnerCFL}\ee For a non zero strange quark
mass the integrals in Eqs. (\ref{self-energy}) and (\ref{tadpole})
have to be evaluated numerically.  In Fig. \ref{masse12and38} we
present the results for the squared Meissner masses of gluons with
color $a=1,2,3,8$ in units of $m^2_{M}$. The solid line denotes
gluons with color $a=1,2$; the dashed line gluons with color $a=3$;
finally the dot-dashed line gluons with $a=8$ (the physical eighth
gluon is obtained by a mixing with the photon
\cite{Litim:2001mv,Schmitt:2003aa} and its mass is only proportional
to the mass of the unrotated gluon; however in the ratio
$m^2_M(M_s)/m^2_M(0)$ the proportionality constant cancels out). We
find that increasing $M_s^2/\mu_b$ the degeneracy in the Meissner
masses is partially removed. Moreover there is a discontinuity of
the squared Meissner mass of gluons of colors $a=1,2,3,8$
\cite{ringrFuku,Alford:2005qw}, which at the onset of the gCFL
phase, i.e. for $M_s^2/\mu_b \sim 2\Delta$, drop to negative values.
Thus we find an instability in the gCFL phase analogous to the one
observed by Huang and Shovkovy \cite{Huang:2004bg} in the g2SC case.

\begin{figure}[t]
\centering {\includegraphics[width=14cm]{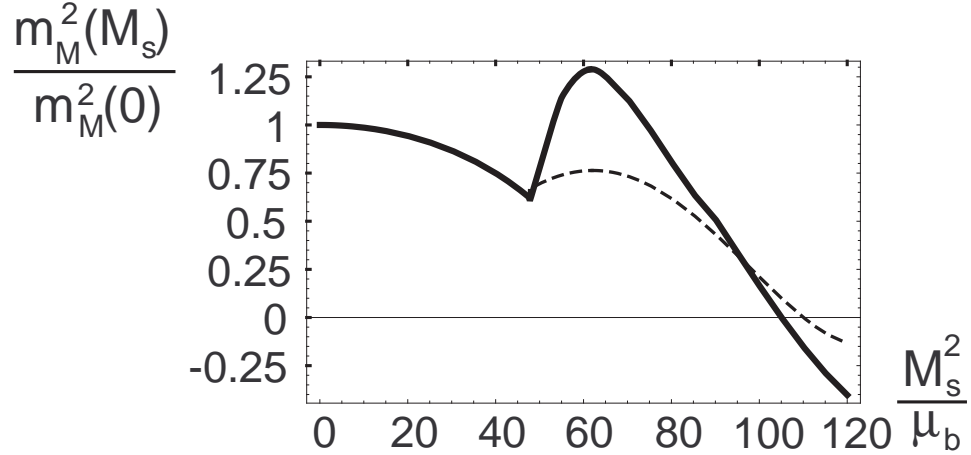}}
\caption{\label{masse45and67}{ \rm Squared values of the  Meissner
masses in units of $m_M^2$ as a function of $M_s^2/\mu_b$. Dashed
line denotes the gluons with colors $a=4,5$; solid line the gluons
with colors $a=6,7$. }}
\end{figure}

In Fig. \ref{masse45and67} we present the results for the gluons
with color $a=4,5$ (solid line) and  color $a=6,7$ (dashed line).
Also in these cases the squared Meissner mass of  gluons are
continuous functions of $M_s^2/\mu_b$. One can notice that, for
very large values of the strange quark mass, the squared Meissner
masses of these gluons become negative. However this  result is
not robust because in the computation of the polarization tensor
Eq.(\ref{polarization}) we have discarded terms  of order ${\cal
O}(M_s/\mu_b)^2$. Therefore, to establish the instability related
to the gluons $a=4,5,6$ and 7 a more accurate analysis would be
needed. Also in this case, as with previous Fig. 1, our results
give not only the Meissner masses in the gCFL phase, but also
their dependence on the strange quark mass in the CFL phase, i.e.
for $M_s^2/\mu_b\,\le\, 2\Delta$.

The instability we have found means that the vacuum was not
correctly identified.  A possible origin of the instability is a non
vanishing vacuum expectation value (vev) of one (or more) time
components of the gluon operator $A^\mu_a$: $<A^0_a>\neq 0$ (see
\cite{Gerhold:2003js}). Clearly defining a new field operator with
vanishing vev's adds contributions that, for $a=3,8$, act as
effective chemical potential terms in the lagrangian: $\sim g<A^0_a>
\psi^\dagger \lambda_a\psi$. The presence of these new terms would
alter the previous results and may lead to real Meissner masses. We
have numerically checked that, either with $<A^0_3>\neq 0$ and the
other vev's $<A^0_a>= 0$, or with $<A^0_8>\neq 0$ and the other
vev's equal to zero, one removes the instability (numerically the
non vanishing vev's must be of the  order of $\sim 10$ MeV). At
present the physical mechanism at the basis of this gluon
condensation is still unclear and we do not push the analysis any
further since our purpose here
 is  to indicate the instability in gCFL
 and not to fully discuss its
antidotes (for  possibly relevant discussion see
\cite{Dietrich:2003nu}). In any case, given the instability of the
gCFL phase, other patterns of condensation, e.g. spin-one color
superconductivity, should be also considered (for a recent
analysis see \cite{Schmitt:2003xq} and references therein).
\section{Conclusions \label{conclusions}}

It is well established that at asymptotic large densities quark matter
is in the CFL phase. At lower densities, in a range presumably more relevant
for the  study of compact stars, neutrality, together with finite strange
quark mass, suggests the gCFL (gapless CFL) phase as the next occurring
ground state. Our calculations in this note suggest an instability of the
gCFL phase, a phenomenon analogous to what observed in the two flavor case.
The instability arises because gluons of color indices $1$, $2$, $3$ and $8$
present an imaginary mass. Its removal may require a different
condensation pattern, most probably including gluon condensation.

\section*{Acknowledgements}
We thank M.~Ciminale for comments. One of us (M.M.) would like to thank
R. Rapp for useful discussions.

%\bibliographystyle{unsrt}
%\bibliography{gCFL8sett}
\end{document}